# Interband Tunneling for Hole Injection in III-Nitride Ultraviolet Emitters

Yuewei Zhang,[1,a] Sriram Krishnamoorthy,[1] Jared M. Johnson,[3] Fatih Akyol,[1] Andrew Allerman,[2] Michael W. Moseley,[2] Andrew Armstrong,[2] Jinwoo Hwang,[3] and Siddharth Rajan[1,3,a]

[1] Department of Electrical and Computer Engineering, The Ohio State University, Columbus, Ohio, 43210, USA

[2] Sandia National Laboratories, Albuquerque, New Mexico 87185, USA

[3] Department of Materials Science and Engineering, The Ohio State University, Columbus, Ohio, 43210, USA

**Abstract:** Low p-type conductivity and high contact resistance remain a critical problem in wide band gap AlGaN-based ultraviolet light emitters due to the high acceptor ionization energy. In this work, interband tunneling is demonstrated for non-equilibrium injection of holes through the use of ultra-thin polarization-engineered layers that enhance tunneling probability by several orders of magnitude over a PN homojunction. $Al_{0.3}Ga_{0.7}N$ interband tunnel junctions with a low resistance of $5.6 \times 10^{-4}$ $\Omega$ $cm^2$ were obtained and integrated on ultraviolet light emitting diodes. Tunnel injection of holes was used to realize GaN-free ultraviolet light emitters with bottom and top n-type $Al_{0.3}Ga_{0.7}N$ contacts. At an emission wavelength of 327 nm, stable output power of 6 $W/cm^2$ at a current density of 120 $A/cm^2$ with a forward voltage of 5.9 V was achieved. This demonstration of efficient interband tunneling introduces a new paradigm for design of ultraviolet light emitting diodes and diode lasers, and could enable higher efficiency and lower cost ultraviolet emitters.

a) Authors to whom correspondence should be addressed.
   Electronic mail: zhang.3789@osu.edu, rajan@ece.osu.edu

AlGaN based ultraviolet light emitting diodes (UV LEDs) are promising for applications in multiple areas, including water purification, biological analysis, sensing, and epoxy curing.[1] In the past decade, research on AlGaN UV LEDs have led to demonstration of emitters over a range of UV spectrum wavelengths.[2-7] However, the wall plug efficiency (WPE) of UV LEDs remains orders of magnitude lower than that of InGaN based visible LEDs[6,8] even though high internal quantum efficiency (IQE) has been achieved in AlGaN quantum wells (QWs) (> 50%)[9,10]. The main reason for the low wall plug efficiency is the poor hole doping and transport in AlGaN. The high acceptor energy (~630 meV in AlN[5]) leads to losses in both injection efficiency (due to low hole density) and electrical efficiency (due to high p-type contact and specific resistance).

Previous designs have used two approaches to overcome the p-type resistance issue. The first involves using a thick p-GaN contact region[3,4,6] to improve hole injection. This approach does enable relatively low electrical losses, but leads to absorption losses in the p-GaN region, leading to high extraction losses. The second approach uses a transparent p-AlGaN layer or an AlGaN/GaN superlatice to inject holes[2,5,6]. While this approach reduces absorption and extraction losses, leading to improved EQE, but the resistive losses in the top AlGaN are high, leading to high electrical losses and low wall-plug efficiency. Both approaches to solve the hole conductivity problem therefore lead to a reduction in the wall plug efficiency.

The approach (Figure 1) we adopt in this work is to use polarization-engineered tunnel junctions[11-15] integrated on UV emitters to reduce both absorption and electrical losses. Using low-resistivity tunnel junctions and thin p-regions minimizes the electrical losses for hole injection. Furthermore, the hole availability is not limited by thermal ionization of acceptors since holes can be introduced through *non-equilibrium injection* across the tunnel junction. This can be especially useful for high Al-composition structures where the equilibrium hole concentration is extremely low. The increased availability of holes could have direct impact on the external quantum efficiency. In addition, the top of the structure is n-type AlGaN, which is transparent and has low spreading resistance. Since top contacts with low metal coverage can be used, light can be extracted directly from the top side without any need for flip chip bonding.[14,17,18] The critical component of such an approach is the tunnel junction, which becomes increasingly challenging as the band gap is increased. Previously, it was shown that GaN tunnel junctions (band gap ~ 3.4 eV) with a low resistance of ~ $1\times10^{-4}$ Ohm cm$^2$ can be achieved using a polarization engineered approach.[15] In this work, we show that it is possible to use polarization engineering to realize interband tunnel junctions in materials with UV-relevant band gaps greater than 4 eV, and report resistance lower than $5.6\times10^{-4}$ Ohm cm$^2$ for $Al_{0.3}Ga_{0.7}N$ tunnel junctions.

The epitaxial structure and energy band diagram investigated in this work are shown in Figure 2. The device consists of a PN junction UV LED with $Al_{0.2}Ga_{0.8}N$ quantum wells, capped by a p-$Al_{0.3}Ga_{0.7}N$/ $In_{0.25}Ga_{0.75}N$/ n-$Al_{0.3}Ga_{0.7}N$ backward tunnel diode. The TJ-based UV LED structure as shown in Fig. 2(b) contains a UV LED structure, a TJ layer, and a top n-AlGaN transparent contact layer. The structure was grown by N$_2$ plasma assisted molecular beam epitaxy (MBE) on Si-doped $Al_{0.3}Ga_{0.7}N$ template, which was grown by metalorganic chemical vapor deposition (MOCVD) with a threading dislocation density of ~ $1.5\times10^{9}$ cm$^{-2}$. Three 4.5 nm $Al_{0.2}Ga_{0.8}N$ QWs were inserted between n and p cladding layers with a 12 nm p type $Al_{0.46}Ga_{0.54}N$ electron blocking layer (EBL) above the quantum wells. A p+ AlGaN / 4 nm $In_{0.25}Ga_{0.75}N$/ graded n+ $Al_{0.3}Ga_{0.7}N$ TJ layer was grown for tunneling contact and hole injection. The InGaN layer growth condition was calibrated by X-ray diffraction (XRD) fitting of bulk InGaN growth.[19] The n+ AlGaN layer was graded from $Al_{0.22}Ga_{0.78}N$ to $Al_{0.3}Ga_{0.7}N$ in 15 nm, to lower the depletion barrier and avoid light absorption/ re-emission. The high-angle annular dark-field scanning transmission electron microscopy (HAADF-STEM) image of a typical TJ-based UV LED structure (Figure 2(a)) show that flat and sharp heterointerfaces were achieved for the quantum wells and the $In_{0.25}Ga_{0.75}N$ interband tunneling layer.

Inductively coupled plasma reactive ion etching (ICP-RIE) with $BCl_3$/ $Cl_2$ chemistry was used for device mesa isolation. Bottom contact evaporation was done with a metal stack of Ti(20 nm)/ Al(120 nm)/ Ni(30 nm)/ Au(50 nm). Rapid thermal annealing under $N_2$ atmosphere at 850 °C for 30 seconds was then carried out to form ohmic contact to the bottom contact layer. Al(20 nm)/ Ni (20 nm)/ Au(80 nm) was then deposited for top contact. Full metal coverage was used for measuring the electrical behavior, while partial metal coverage was used for optical power measurements. For 50 × 50 µm$^2$ devices, ~ 28% of the device region is covered with L shape metal to minimize the blocked light intensity. The electroluminescence and emission power were obtained from on-wafer measurement at room temperature using a calibrated Ocean Optics USB 2000 spectrometer with a coupled fiber optic cable.

Fig. 2(c) shows the band diagram of the TJ UV LED structure calculated using a one dimensional Schroedinger-Poisson solver.[20] A TJ layer on top of p-AlGaN layer enables tunneling contact. The high polarization charge density at the AlGaN/ InGaN interface builds up high polarization fields, causing band bending across the thin InGaN layer to align the band edges of n+ and p+ AlGaN over just a few nanometers.[15,21] When the LED is forward biased, the top TJ layer is reverse biased, electrons tunnel from the valence band in p-AlGaN to the conduction band in n-AlGaN, and the remnant holes are then injected into p-AlGaN. As shown in Fig. 2(c), the tunnel barrier consists of the (interband) tunnel barrier across the InGaN bandgap and (intraband) depletion barriers in the p+ and n+ AlGaN layers. Since the depletion barrier in the n side is higher, a graded n+ AlGaN layer was used to reduce the barrier height and increase effective tunneling probability.

The electrical characteristics of devices (50 × 50 µm$^2$) with full and partial top metal coverage are shown in Figure 3(a). The series resistance of the device is not constant over the entire current range but decreases as the current density is increased, with non-linearity originating both from the LED, and the tunnel junction. At a current density of 2 kA/cm$^2$, the voltage drop was 7.47 V, while the voltage drop at 20 A/cm$^2$ was 4.8 V. In the case of the device with partial metal coverage, we observed a slightly higher voltage drop, which we attribute to the spreading resistance in the n-type AlGaN. Increasing the thickness of the Al$_{0.3}$Ga$_{0.7}$N top contact layer would reduce the spreading resistance.

The differential resistance of the device is shown in Fig. 3(b). The resistance decreases with increasing current density, and reaches a minimum of 7.5×10$^{-4}$ Ohm cm$^2$ above 1 kA/cm$^2$. To estimate the contribution of the tunnel junction to this total resistance, we de-embedded the other various components in the resistance. The contact resistance for the top and bottom regions were estimated (from transfer length method measurements) to be ~ 1.4×10$^{-6}$ Ohm cm$^2$ and ~ 4.6×10$^{-6}$ Ohm cm$^2$, respectively, and the series resistance of the p-AlGaN layer was estimated to be ~ 1.9×10$^{-4}$ Ohm cm$^2$ (acceptor activation energy = 220 meV[22,23], and hole mobility = 1 cm$^2$/Vs[5]). The resistance of n-AlGaN layers is ignored. The resistance of the p+ AlGaN/ InGaN/ n+ AlGaN TJ layer can be estimated as 5.6 ×10$^{-4}$ Ohm cm$^2$. Fig. 4 shows previous reports of tunnel junction resistance as a function of band gap, together with the present result. As expected from tunneling theory, the resistance increases exponentially as the barrier for interband tunneling increases, and the reported values for homojunctions fall along a fairly consistent trendline. The use of polarization-engineered tunnel junctions can therefore enable low-resistance tunnel junctions that are many orders of magnitude lower than would be possible with a PN junction tunnel junction.

To understand the dependence of tunneling resistance on current, we estimated the tunneling current using a semi-classical approximation[34]:

$$J_T = \frac{4\pi q m^*}{(2\pi)^3 \hbar^3} \int_0^\infty dE_z T(E_z) \int_0^\infty dE_t \left( f_n^p(E_z, E_t) - f_n^n(E_z, E_t) \right)$$

where $m^*$ is the effective mass, $E_z$ and $E_t$ are the transverse and perpendicular kinetic energies, $f_n^p$ and $f_n^n$ are the Fermi-Dirac distributions of electrons in the p+ and n+ AlGaN layers respectively. $T(E_z)$ represents the tunneling probability for an electron with a z-directed kinetic energy $E_z$[13,15] and was evaluated using the Wentzel–Kramers–Brillouin (WKB) approximation. The potential profile for the calculation was extracted from self-consistent Schrodinger-Poisson calculations.[15] The dependence of TJ resistance on current density is shown in Fig. 3(b) (dashed line) for the TJ design used here, and is in good agreement with experimental results.

Electroluminescence (EL) characteristics of the TJ UV LED structures were measured to confirm hole injection. Both top and bottom contacts are made to n-type AlGaN. As shown in Figure 5, single peak emission at 327 nm was obtained with a full width at half maximum of around 14.7 nm, and no significant long wavelength peaks were observed. A small blue shift with increasing current due to the quantum confined Stark effect is evident.[35] The inset to Fig. 5 shows an optical micrograph of a TJ UV LED device showing a 50 × 50 μm² mesa with top contact metal along the borders (dark L-shape), and no metal in the remaining region. When operated at 10 mA, emission is evident over the entire mesa showing that the LED operates in regions without any metal coverage since the n-AlGaN is efficient at spreading the current.

The emission power spectrum as a function of injected current (continuous wave) is shown in Figure 6(a). The measured output power of 0.58 mW at 20 mA, corresponds to a normalized optical power of 23 W/cm$^2$ at 800 A/cm$^2$. The maximum external quantum efficiency and wall plug efficiency of the sample are 1.5% and 1.08% as shown in Fig. 6(b) and (c), respectively. The measured power underestimates the actual power output because the LED was measured on-wafer without an integrating sphere, and with 28% of the device covered with a thick absorbing metal contact. The output power would be higher in a packaged device with minimized metal coverage.

While the output power and efficiency reported here are likely limited by the internal quantum efficiency due to the high dislocation density ($> 10^9$ cm$^{-2}$) substrate and unoptimized active region design, the results demonstrate the potential for tunnel junctions to enable highly efficient hole injection with low electrical and absorption losses. The work reported here could lead to several new avenues for research for UV emitters. With tunneling-based non-equilibrium hole injection, the efficiency of UV LEDs will be impacted less by the low hole density, especially in higher AlGaN composition devices. This could greatly improve the injection efficiency and reduce electron overflow. Since the tunneling resistance is low, and since n-type contacts are usually much less resistive than p-type contacts, the electrical losses are reduced. The use of tunnel junctions enables several new fabrication and device innovations, such as roughening of the epitaxial top surface to enable better light extraction, multiple active region LEDs to increase output power and reduce the cost, and multiple active region devices to emit different wavelengths for broadband or multi-color sources. Finally, polarization-engineered efficient tunnel injection of carriers into wide bandgap materials could be extended to other wide band gap materials where complimentary doping is difficult, such as ZnO.

One drawback of the approach described here is the absorption in the InGaN layer used in the tunnel junction. However, the actual losses realized due to this are low since the layer is very thin, and because the high field in the InGaN layer sweeps any photogenerated carriers back into the active region. Assuming an absorption coefficient ($\alpha$) of $1\times10^5$ cm$^{-1}$ [36], the fraction of power absorbed in 4 nm thick InGaN is less than 4%. Since the field in this layer is high, the holes are swept back into the p-AlGaN layer. We estimate (Supplementary Information) that for an absorbing layer with absorption $L_A$, and an active region with internal quantum efficiency $R$, the maximum actual loss is given by $L_A(\frac{1-R}{1-L_A R})$. For an internal quantum efficiency of 50% (which is now achievable in UV LEDs), the effective loss is therefore less than 2%.

In conclusion, we have demonstrated non-equilibrium hole injection into the active regions of the UV LED structures through tunnel junction. Tunneling resistance of $5.6\times10^{-4}$ Ohm cm$^2$ was obtained for an Al$_{0.3}$Ga$_{0.7}$N tunnel junction. Hole injection was confirmed through electroluminescence measurements of 327 nm LEDs integrated with the tunnel junctions. The demonstration of tunneling in a wide band gap (> 4 eV) material shows that nanoscale heterostructure engineering can overcome limits on traditional tunnel junctions, and could enable a new generation of higher efficiency and low cost ultraviolet solid state light sources.

Acknowledgement:

S.R., Y.Z., S.K. and F.A. acknowledge funding from the National Science Foundation (ECCS-1408416). Sandia National Laboratories is a multi-program laboratory managed and operated by Sandia Corporation, a wholly owned subsidiary of Lockheed Martin Corporation, for the United

Figure Captions:

Fig. 1 Schematic design of a TJ-based UV LED with tunnel junctions integrated on a UV emitter to reduce both absorption and electrical losses.

Fig. 2 (a) HAADF-STEM image, (b) epitaxial stack, and (c) equilibrium energy band diagram of TJ-based UV LED structure.

Fig. 3 (a) Electrical characteristics of TJ UV LED devices ($50 \times 50$ μm$^2$) with full and partial top metal coverage. (b) Differential resistance as a function of current density (full top metal coverage).

Fig. 4 Reported TJ resistance as a function of band gap energy for different material systems. The tunnel junction reported here is for $Al_{0.3}Ga_{0.7}N$ (band gap ~ 4.3 eV) with resistance of $5.6 \times 10^{-4}$ Ohm cm$^2$.

Fig. 5 Electroluminescence of the TJ UV LED structure with dc current injection from 0.1 mA to 20 mA at room temperature, single peak emission at 327 nm is shown. The inset shows an optical micrograph of a TJ UV LED device ($50 \times 50$ μm$^2$) with partial top metal coverage operated at 10 mA.

Fig. 6 (a) Output power, (b) EQE and (c) WPE of the $50 \times 50$ μm$^2$ TJ UV LED device. The powers were measured on wafer without integrating sphere from the top surface.

## TJ UVLED

| n AlGaN |
| Tunnel Junction |
| Thin p AlGaN |
| MQW |
| n AlGaN |

LED

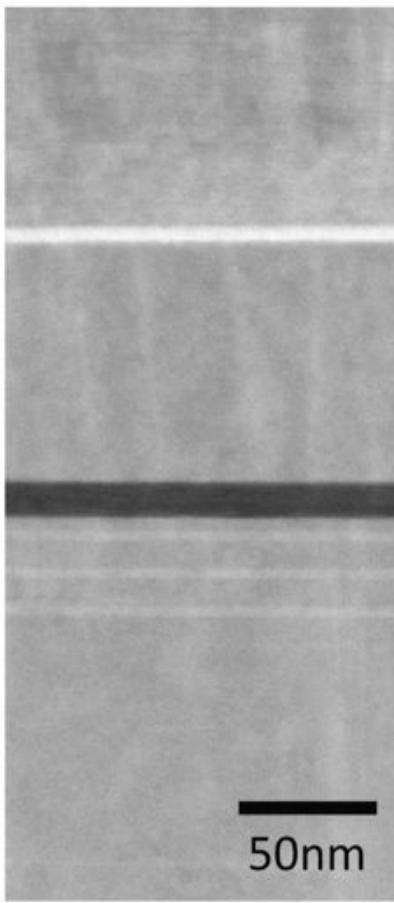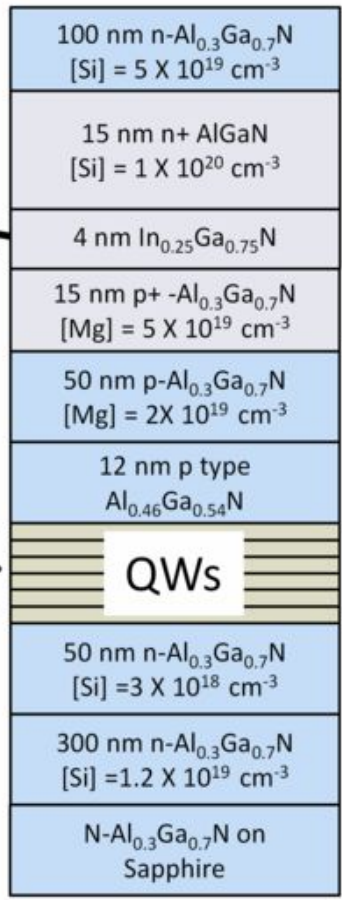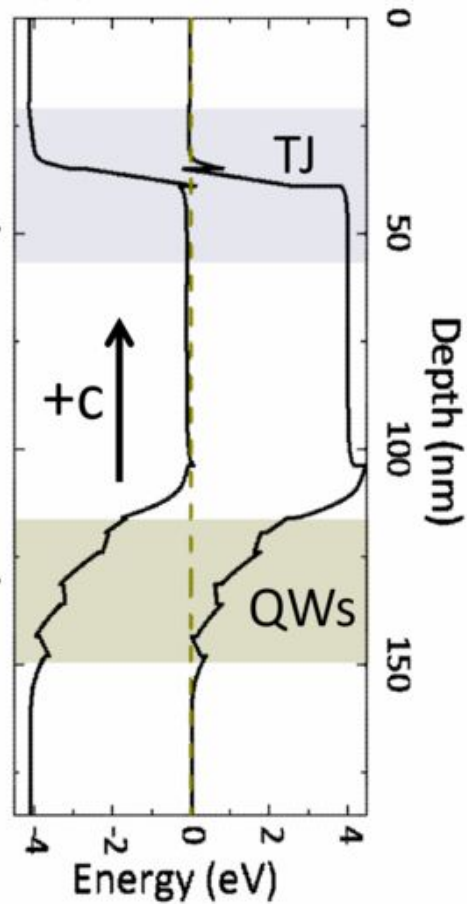

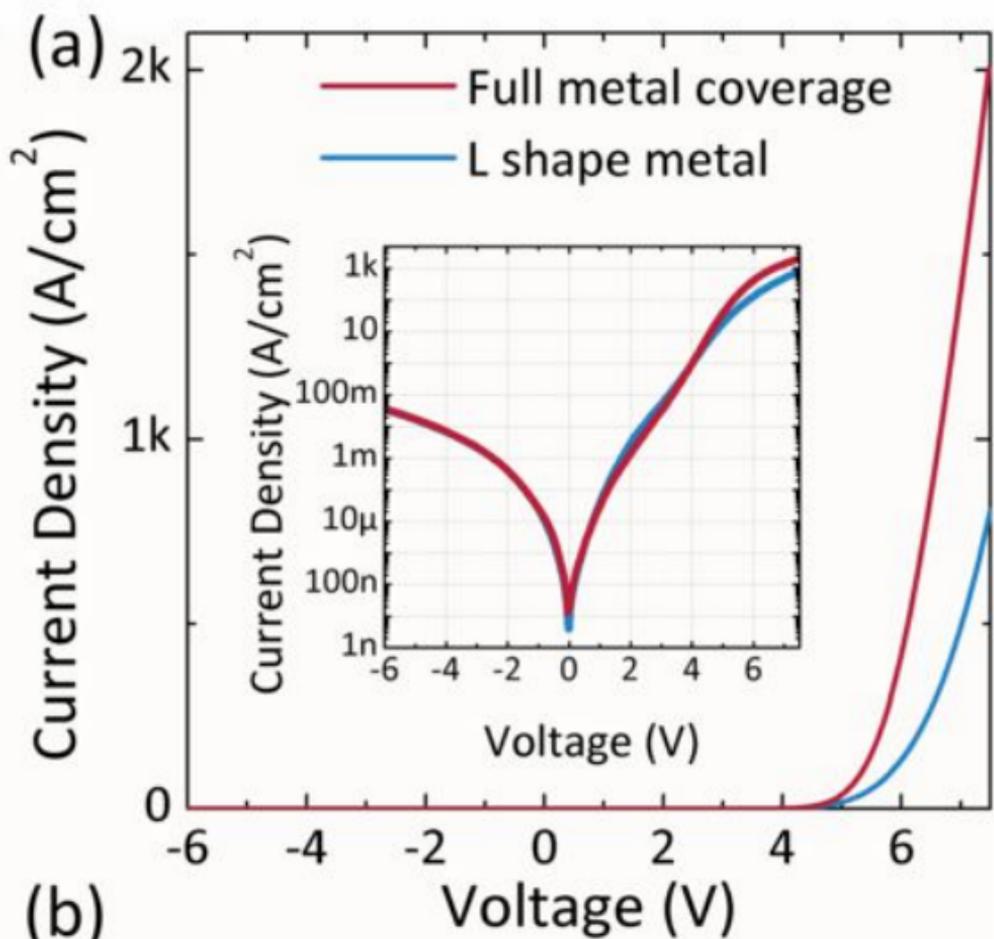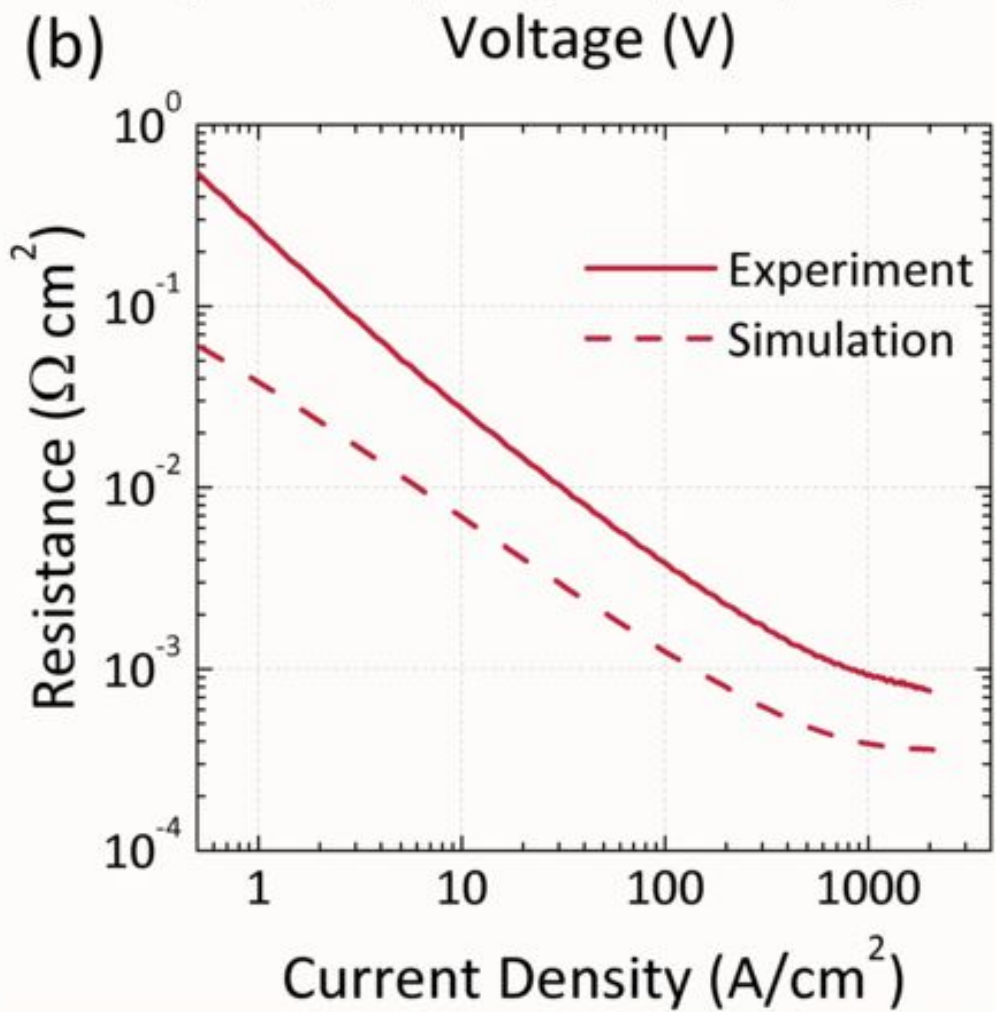

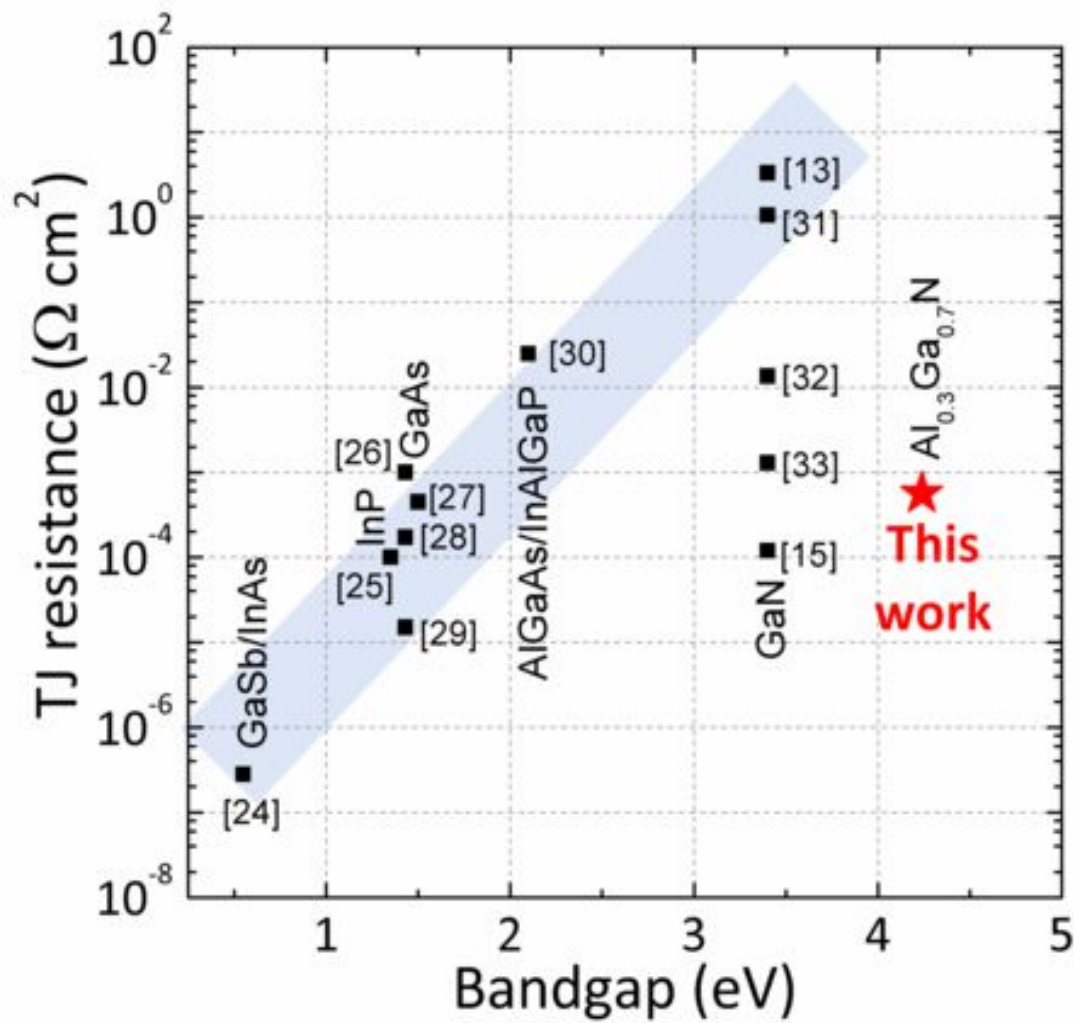

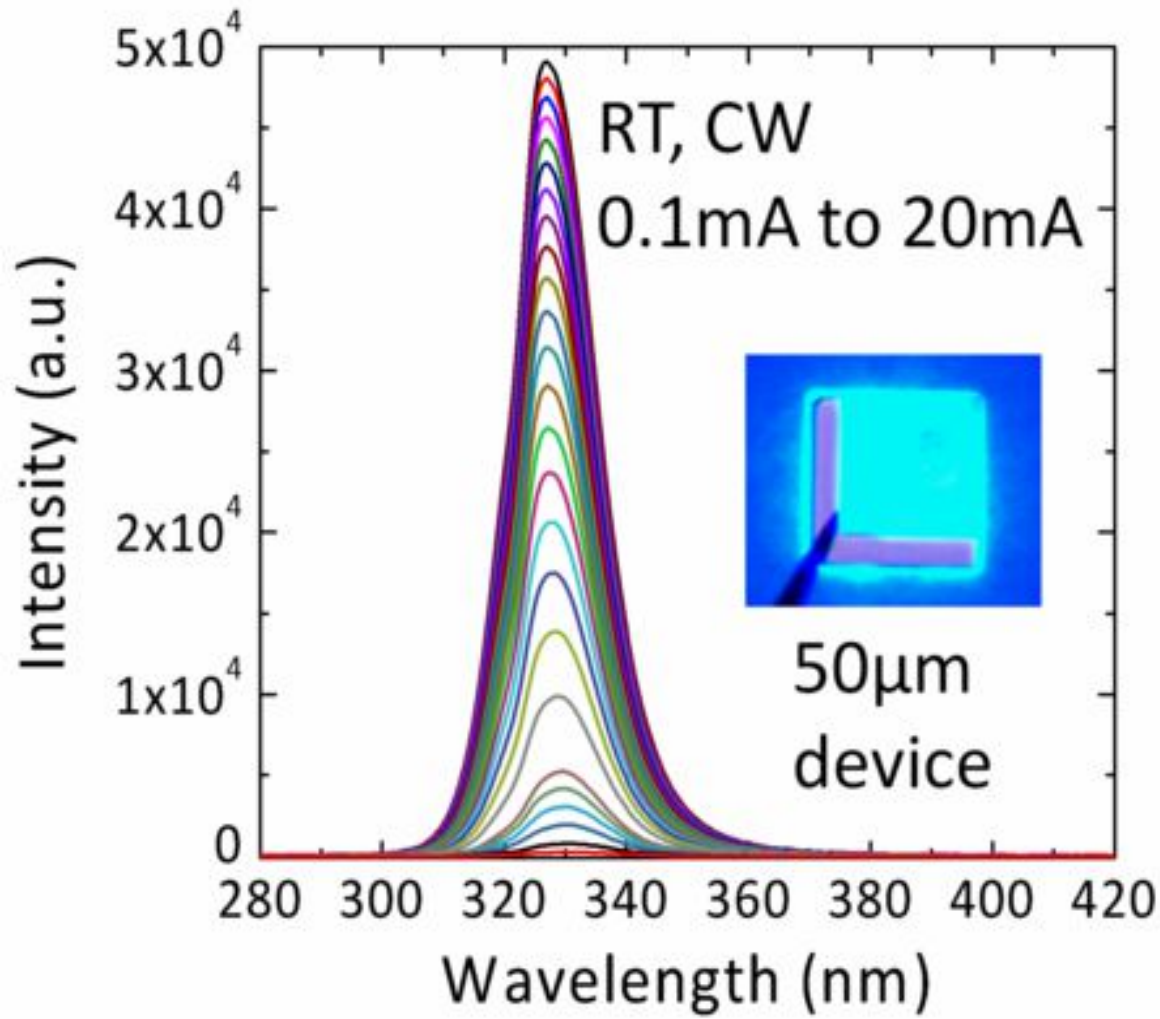

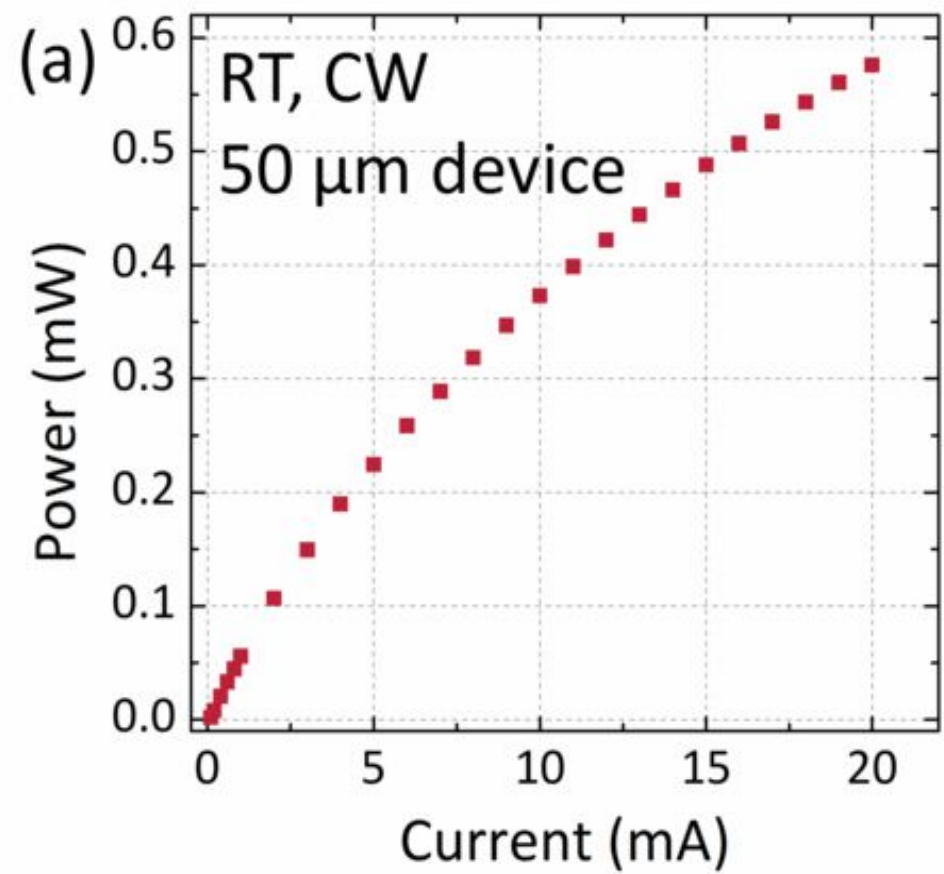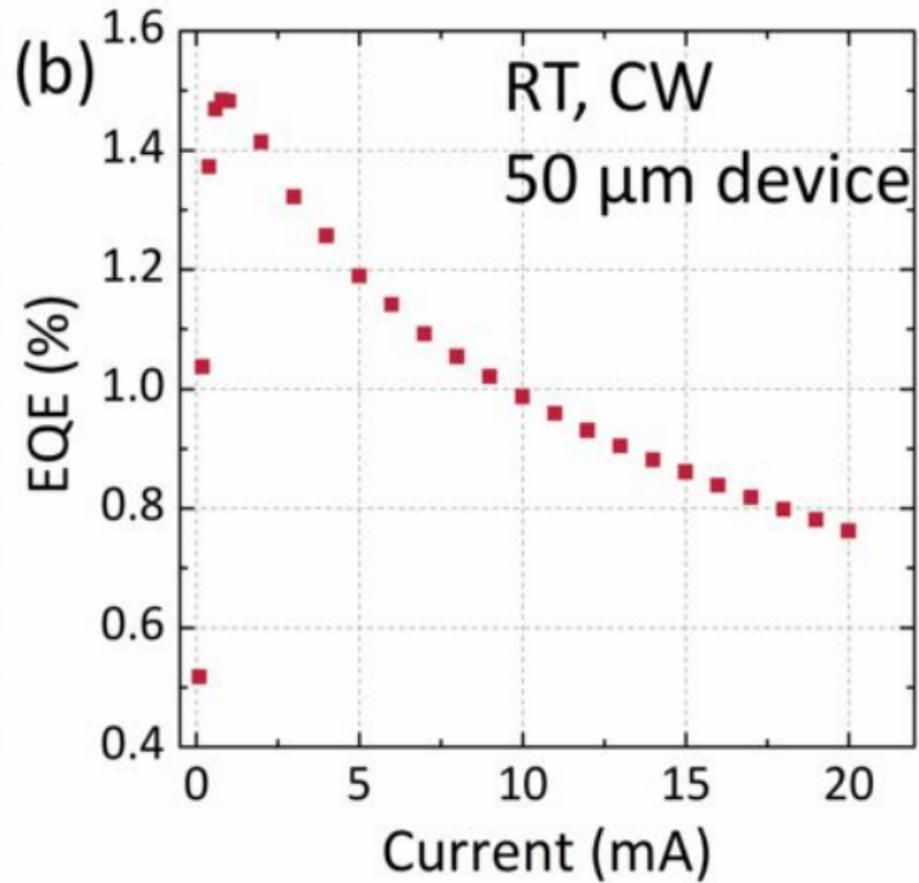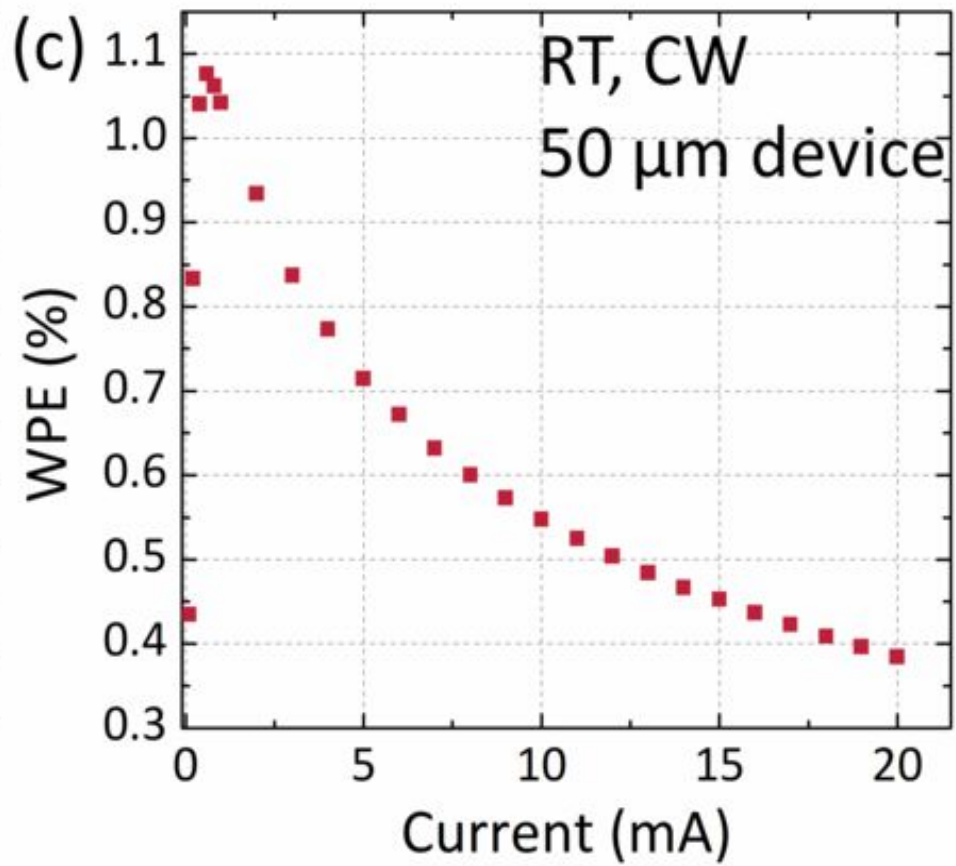